\documentclass[a4paper,11pt]{article}
\pdfoutput=1 
\usepackage[utf8]{inputenc}
\usepackage{jheppub} 
\usepackage[T1]{fontenc} 
\usepackage{slashed}
\usepackage{subcaption}
\usepackage{url}
\newcommand{\ifb}{\text{fb}^{-1}}
\newcommand{\bea}{\begin{eqnarray}}
\newcommand{\eea}{\end{eqnarray}}

\title{\boldmath 
Electroweak Loops as a Probe of New Physics in $t\bar{t}$ Production at the LHC
}

\author{Till Martini}
\author{and Markus Schulze}
\affiliation{Humboldt-Universit\"at zu Berlin,\\Institut f\"ur Physik,\\Newtonstrasse 15,\\12489 Berlin, Germany}

\emailAdd{till.martini@physik.hu-berlin.de}
\emailAdd{markus.schulze@physik.hu-berlin.de}

\abstract{
We calculate the $\mathcal{O}(\alpha)$ weak corrections to top quark pair production at the LHC 
and include anomalous electroweak interactions from dimension-six operators.
The loop calculation and renormalization are consistently done  within the Standard Model Effective Field Theory.
Sensitivity to the involved operators is exposed through the virtual corrections, which receive enhancement from electroweak Sudakov logarithms.
We investigate the prospects of using this feature for probing New Physics at the LHC that so far has only been studied in final states with on-shell sensitivity such as $t\bar{t}+Z$ or $t \to b W$.
We find that the large $t\bar{t}$ production rate and the excellent perturbative control allow compensating the loop suppression and
yield remarkably strong constraints that are competitive with those from $t\bar{t}+Z$. 
}

\begin{document} 
\hfill{\vspace{-3ex}HU-EP-19/36}\\
\maketitle
\flushbottom

\section{Introduction}
Run-2 data taking of the Large Hadron Collider (LHC) lasted from early 2015 to mid 2018 and culminated in a data set of about $160~\ifb$ integrated luminosity
at 13~TeV collision energy. 
In this period the top quark was produced approximately 240 millions times.
This is more than a thousand times more frequent than it was ever produced at its discovery machine -- the Tevatron.
Soon, around the year 2023, the integrated LHC luminosity is expected to have doubled reaching up to $400~\ifb$.
The wealth of this experimental data yields unprecedented precision in hadron collider physics 
and enables powerful probes of New Physics in the top quark sector, if combined with precise theory predictions. 

Fortunately, top quark production and decay at the LHC is theoretically under very good control:
Next-to-next-to-leading (NNLO) order in Quantum Chromodynamics (QCD) predictions are available for both, single and pair production~\cite{Czakon:2013goa,Brucherseifer:2014ama,Czakon:2015owf,Czakon:2016ckf}.
The decay dynamics are known to the same order~\cite{Czarnecki:1998qc,Chetyrkin:1999ju,Gao:2012ja,Brucherseifer:2013iv,Czarnecki:2018vwh}, and have been combined with the production processes
via the narrow width approximation in Ref.~\cite{Berger:2016oht,Berger:2017zof,Gao:2017goi,Behring:2019iiv}.
Electroweak corrections are also known for the production process~\cite{Beenakker:1993yr,Kuhn:2005it,Bernreuther:2005is,Bernreuther:2006vg,Moretti:2006nf,Kuhn:2006vh,Kuhn:2013zoa,Czakon:2017wor} and the top quark width~\cite{Do:2002ky,Basso:2015gca}. 
QCD resummation has been considered in Refs.~\cite{Beneke:2009ye,Czakon:2009zw,Beneke:2011mq,Cacciari:2011hy,Kidonakis:2012rm,Ferroglia:2012ku,Ferroglia:2013awa,Czakon:2018nun} up to next-to-next-to-leading logarithmic (NNLL) level, and 
studies that go beyond the narrow width approximation have been presented in Refs.~\cite{Bevilacqua:2010qb,Denner:2012yc,Heinrich:2013qaa,Frederix:2016rdc,Denner:2016jyo,Jezo:2016ujg} through higher orders.
Taken all together, a picture arises in which the main top quark dynamics can be predicted at the few percent level,
from threshold up to highest energies.

In this paper, we ask the question whether one can exploit the high precision in the top quark sector to 
probe New Physics that enters through {\it virtual loop corrections}.  
Historically, this is a very fruitful approach as, for example, the top quark was postulated by considering kaon mixing through loops, 
decades before its discovery~\cite{Kobayashi:1973fv,Harari:1975aw}. 
Also the Higgs boson mass was indicated accurately  by LEP precision fits arising from electroweak loops~\cite{Barate:2003sz}. 
Similarly, many New Physics searches at $B$-factories harvest loop sensitivity and place incredibly strong limits on deviations from the Standard Model (see e.g., Ref.~\cite{Kou:2018nap}).
Top quark pair production at the LHC might be a similarly powerful probe.
Besides resonance searches in the $t\bar{t}$ mass spectrum and modifications of QCD interactions, the most 
interesting corner to look for New Physics is the electroweak top quark interactions.
Prominent examples are the analyses of $t\bar{t}+Z/W/H$~\cite{Sirunyan:2017uzs,Sirunyan:2018koj,CMS:2019too,Aaboud:2019njj} final states or the $W$-helicity fractions in top quark decays~\cite{Aaboud:2016hsq,Khachatryan:2016fky}, which 
yield direct {\it on-shell} sensitivity to the couplings between top quarks and the electroweak bosons. Global fits of New Physics contributions in terms of dimension-six operators relevant for the top quark sector have been carried out in Refs~\cite{Hartland:2019bjb,Brivio:2019ius,Durieux:2019rbz}.
The resulting constraints on New Physics are substantial (see e.g. Ref.~\cite{CMS:2019too}), but the above analyses have their limitations:
1) The associated production processes $t\bar{t}+X$ suffer from relatively small cross sections because of coupling suppression, a high production threshold and penalties from branching fractions;
2) Sensitivity in top quark decay dynamics is often diminished by cancellations between the squared matrix element in the numerator and the total width in the denominator. 
It is therefore essential to explore alternative avenues such as sensitivity to New Physics from electroweak loop corrections. 
Let us substantiate this reasoning by presenting a first estimate of such prospects. 
We compare the tree level $pp \to t\bar{t}+Z$ process at $\mathcal{O}(\alpha_s^2 \alpha)$ with the 
electroweak correction to the $pp \to t\bar{t}$ process, also at $\mathcal{O}(\alpha_s^2 \alpha)$.
For a final state with a leptonic $Z$ boson decay and semi hadronic top quark decays, one finds
\bea
\label{eq:intro1}
  \sigma_{t\bar{t}Z} \times \mathcal{B}_{Z\to\ell\ell} \times \mathcal{B}_{t\bar{t}\to \ell\nu+\mathrm{jets}} \approx 1\;\mathrm{pb} \times 6\% \times 33\% \approx 20\;\mathrm{fb}
\eea
for the $t\bar{t}Z$ process~\cite{Tanabashi:2018oca} and for the next-to-leading order (NLO) electroweak contribution to $t\bar{t}$ production%
\footnote{We neglect electroweak Sudakov enhancement~\cite{Sudakov:1954sw,Ciafaloni:1998xg,Kuhn:1999de,Beenakker:2000kb} in this estimate and simply use a multiplication with the fine structure constant $\alpha$.}%
~\cite{Tanabashi:2018oca}
\bea 
\label{eq:intro2}
  \alpha \times \sigma_{t\bar{t}} \times\mathcal{B}_{t\bar{t}\to \ell\nu+\mathrm{jets}} \approx 1/128 \times 840\;\mathrm{pb} \times 33\% \approx 1800\;\mathrm{fb}.
\eea
The striking difference between $t\bar{t}Z$ and $t\bar{t}$ production mainly arises because of the factor $\mathcal{B}_{Z\to\ell\ell}$ that is missing for the virtual corrections in Eq.~(\ref{eq:intro2}). 
Hence, the study of New Physics in electroweak corrections to $t\bar{t}$ seems very promising and appears competitive to similar studies in $t\bar{t}+Z$. 
This rationale is of course very dependent on the respective backgrounds.
While the $pp \to t\bar{t}+Z \to \ell \nu b\bar{b}jj+ \ell\ell$ process has relatively small backgrounds, 
the $t\bar{t}$ final state is dominantly arising from QCD dynamics which act as irreducible background for our purposes. 
Fortunately, the QCD production dynamics are known to very high perturbative order (NNLO+NNLL QCD~\cite{Czakon:2009zw,Beneke:2011mq,Cacciari:2011hy,Czakon:2013goa,Czakon:2018nun}). This should allow for a precise extraction of the electroweak effects which have been found to be comparable or even exceeding the current QCD uncertainties in some phase space regions (in particular for transverse momenta between $500$ GeV and $2$ TeV, see, e.g., Fig.~4 in Ref.~\cite{Czakon:2017wor}).

Building upon this idea, we present a comprehensive investigation in this paper and compare our results with conventional results. 
We calculate the next-to-leading order weak corrections and, for the first time, consistently include the effects of dimension-six operators of a 
Standard Model Effective Field Theory (SMEFT)~\cite{Coleman:1969sm,Callan:1969sn,Weinberg:1980wa,Manohar:1996cq,Grzadkowski:2010es,Willenbrock:2014bja,Passarino:2016pzb,Dedes:2017zog} in this loop calculation. 
A consistent ultraviolet (UV) renormalization within SMEFT guarantees a finite result. 
Our analysis makes use of differential distributions and allows expanding the cross section up to dimension six and dimension eight from squared dimension-six contributions.
We benchmark our results with independent constraints obtained from the $p p \to t\bar{t}+Z$ process.

\section{Calculation}

\subsection{Dimension-six coupling parameterization }

The main aim of this work is the investigation of anomalous top quark electroweak interactions. 
Hence, we focus on the $t\bar{t}Z$ and $t\bar{t}W$ couplings and allow deviations from their SM value within the framework of the Standard Model Effective Field Theory (SMEFT) 
with contributions from dimension-six operators.
In the SM, the $t\bar{t}Z $ vertex is given by 
\bea
\label{eq:ttzvertex}
 \Gamma^\mu_{Ztt} =  \frac{-\mathrm{i}e}{s_\mathrm{w} c_\mathrm{w}} \gamma^\mu 
 \bigg(
   d^Z_\mathrm{L} P_\mathrm{L}
  +
   d^Z_\mathrm{R} P_\mathrm{R}  
 \bigg)
\quad \text{with} \quad
 P_{\mathrm{R/L}} = \frac12 ( 1\pm\gamma_5 )
\eea
and  $ d^Z_\lambda = d^{Z,\mathrm{SM}}_\lambda = T^3_{t_\lambda} - s_\mathrm{w}^2 Q_t  $, where $T^3$, $Q$, and $c_\mathrm{w}=\sqrt{1-s_\mathrm{w}^2}$ are the weak isospin, electric charge, and the cosine of the weak mixing angle, respectively. 
An extension to the $\mathrm{SU(2)\times U(1)}$ symmetric effective field theory in the Warsaw basis~\cite{Grzadkowski:2010es} leads to the following shifts of left-handed and right-handed couplings 
\bea
\label{eq:ttzcoupl}
  d^Z_\mathrm{L} \to d^{Z,\mathrm{SM}}_\mathrm{L}
   + 
   \frac12 \frac{v^2}{\Lambda^2} \left( C_{33}^{\varphi q 3} - C_{33}^{\varphi q 1} \right),
   \quad \text{and} \quad
  d^Z_\mathrm{R} \to d^{Z,\mathrm{SM}}_\mathrm{R}
   -
   \frac12 \frac{v^2}{\Lambda^2}  C_{33}^{\varphi u},
\eea
where $v$ is the vacuum expectation value and $\Lambda$ is the characteristic scale of the Effective Field Theory. 
\begin{figure}[t]
\centering
$\Gamma^\mu_{Ztt} =$
\raisebox{-0.5\totalheight}{\includegraphics[width=0.12\textwidth]{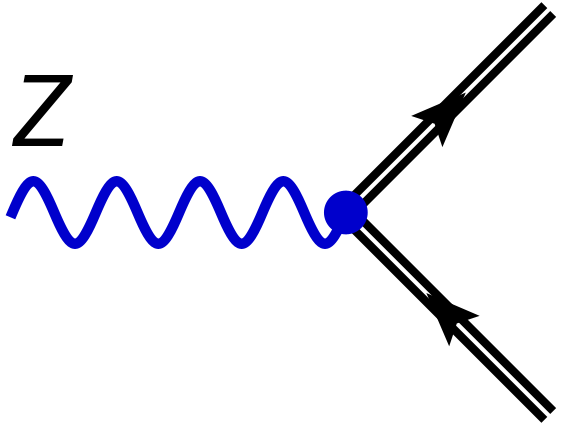}},
\hfill
$\Gamma^\mu_{Wtb} =$
\raisebox{-0.5\totalheight}{\includegraphics[width=0.12\textwidth]{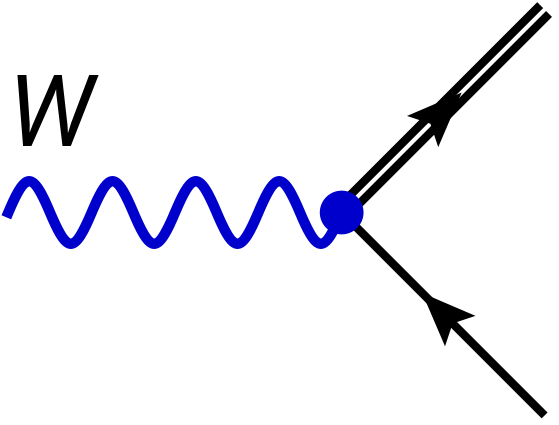}},
\hfill
$\Gamma_{\chi tt} =$
\raisebox{-0.5\totalheight}{\includegraphics[width=0.12\textwidth]{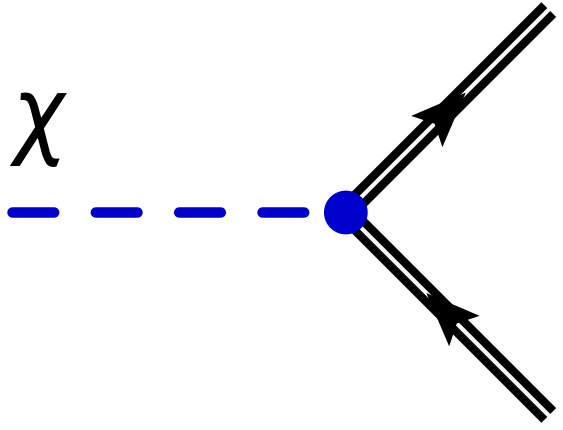}},
\hfill
$\Gamma_{\phi tb} =$
\raisebox{-0.5\totalheight}{\includegraphics[width=0.12\textwidth]{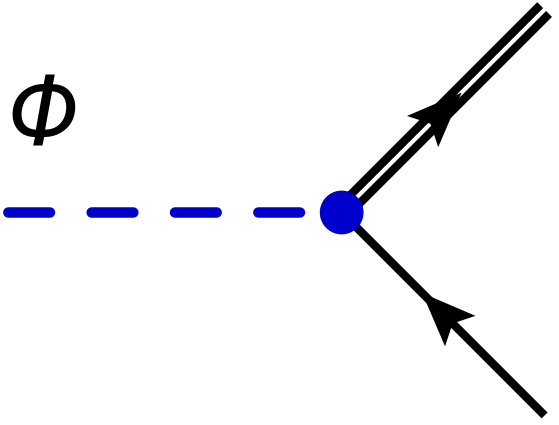}}
\hfill
\caption{\label{fig:i} Electroweak top quark interactions that receive modifications from dimension-six operators in our work. }
\end{figure}
The above Wilson coefficients correspond to the operators 
\bea
\label{eq:EFTop1}
 Q_{33}^{\varphi q 1}  &=& \left( \varphi^\dagger \mathrm{i} \overleftrightarrow{D}_{\hspace{-0.7ex}\mu} \varphi \right) \, ( \bar{t}'_\mathrm{L} \gamma^\mu t'_\mathrm{L} ), \\ 
\label{eq:EFTop2}
 Q_{33}^{\varphi q 3}  &=& \left( \varphi^\dagger \mathrm{i} \tau^I  \overleftrightarrow{D}_{\hspace{-0.7ex}\mu}\varphi \right) \, ( \bar{t}'_\mathrm{L} \tau^I \gamma^\mu t'_\mathrm{L} ), \\
 Q_{33}^{\varphi u} &=& \left( \varphi^\dagger \mathrm{i} \overleftrightarrow{D}_{\hspace{-0.7ex}\mu} \varphi \right) \, ( \bar{t}'_\mathrm{R} \gamma^\mu t'_\mathrm{R} ),
\label{eq:EFTop3} 
\eea
where $\tau^I$ are the Pauli matrices.
For simplicity, we eliminate one of the two coefficients in $d^Z_\mathrm{L}$ of Eq.~(\ref{eq:ttzcoupl}) 
and set $C_{33}^{\varphi q 1} =- C_{33}^{\varphi q 3}$. 
This corresponds to neglecting corrections  to the left-handed $Zb\bar{b}$ vertex that has limited impact on our analysis. As a consequence, shifts in the couplings of $t\bar{t}Z$ and $t\bar{t}W$ are correlated, as are their corresponding would-be Goldstone bosons. 
We also neglect dipole operators and the operator $Q_{\varphi WB}$ \cite{Grzadkowski:2010es} which contributes to $d^Z_\mathrm{L}$. They can potentially be studied in $t\bar{t}\gamma$ final states.
The operator in Eq.~(\ref{eq:EFTop2}) also modifies the $Wtb$ vertex
\bea
\label{eq:tbwvertex}
  \Gamma^\mu_{Wtb} =   \frac{-\mathrm{i} e}{\sqrt{2} s_\mathrm{w}} \gamma^\mu  
  \bigg( d^W_\mathrm{L} P_\mathrm{L} + d^W_\mathrm{R} P_\mathrm{R}  \bigg)
\eea
by shifting the SM contributions $d^{W,\mathrm{SM}}_\mathrm{L}=1-d^{W,\mathrm{SM}}_\mathrm{R}=1$ according to 
\bea
\label{eq:tbwcoupl}
   d^W_\mathrm{L} \to d^{W,\mathrm{SM}}_\mathrm{L} +  \frac{v^2}{\Lambda^2} C^{\varphi q 3}_{33}.
\eea
We assume that there is no contribution to an anomalous right-handed $W$-boson current. 

The vertices connecting the would-be Goldstone bosons $\phi^\pm$ and $\chi^0$ to top (and bottom) quarks receive modifications from the operators in Eqs.~(\ref{eq:EFTop1}--\ref{eq:EFTop3}) as well.
We find 
\bea
\label{eq:topGBvertex}
   \Gamma_{\phi tb} =  \frac{-\mathrm{i}e}{\sqrt{2} s_\mathrm{w} c_\mathrm{w}} 
   \bigg( d^\phi_\mathrm{L} P_\mathrm{L} + d^\phi_\mathrm{R} P_\mathrm{R}  \bigg)
   \quad \text{and} \quad 
   \Gamma_{\chi tt} =  \frac{-e}{2 s_\mathrm{w} c_\mathrm{w}} 
   \bigg( d^\chi_\mathrm{L} P_\mathrm{L} + d^\chi_\mathrm{R} P_\mathrm{R}  \bigg)
\eea
with 
\bea
\label{eq:dGB1}
   d^{\phi^-}_\mathrm{L} & =&   \frac{v^2}{\Lambda^2} \frac{\slashed{p}_{\phi}}{M_Z} C_{33}^{\varphi q3},  \quad\quad
   d^{\phi^+}_\mathrm{L}  =  -\frac{m_t}{M_Z} - \frac{v^2}{\Lambda^2} \frac{\slashed{p}_{\phi}}{M_Z} C_{33}^{\varphi q3},
   \\
\label{eq:dGB2}   
   d^{\phi^-}_\mathrm{R}  & =&  -\frac{m_t}{M_Z}, \quad\quad \quad\quad \quad
   d^{\phi^+}_\mathrm{R}   =  0, 
   \\
\label{eq:dGB3}  
   d^\chi_\mathrm{L} & =& -\frac{m_t}{M_Z} - \frac{v^2}{\Lambda^2} \frac{\slashed{p}_\chi}{M_Z}  \left( C_{33}^{\varphi q 3} - C_{33}^{\varphi q 1} \right),
   \\
\label{eq:dGB4}     
   d^\chi_\mathrm{R} & =& \frac{m_t}{M_Z}  + \frac{v^2}{\Lambda^2} \frac{\slashed{p}_\chi}{M_Z} C^{\varphi u }_{33}.
\eea
The $\slashed{p}$ is the incoming boson momentum. 
In the EFT, there are also four-point vertices $\Gamma_{\phi\phi tt}$ of two would-be Goldstone bosons and two top quarks  
which, in principle, contribute to our process.
However, it turns out that they yield a vanishing contribution after loop integration because their coupling structure 
only yields tensor rank-1 loop integrals of the kind $\int\!\mathrm{d}^D\ell \;\, \ell^\mu \big/ (\ell^2-m^2 +\mathrm{i}\varepsilon) = 0$.

\subsection{$\mathcal{O}(\alpha)$ corrections and UV renormalization in $t\bar{t}$ production}

The tree level amplitudes contributing to $q\bar{q}/gg \to t\bar{t}$ can be categorized into $\mathcal{O}(\alpha_s)$ {\it strong production }
and $\mathcal{O}(\alpha)$ {\it electroweak production}. 
As a result, the pure electroweak $q\bar{q}$ initiated  production cross section is parameterically suppressed by about three orders of magnitude compared to strong production. 
It is further suppressed by the $q\bar{q}$ parton luminosities as compared to gluon fusion, which yields more than $80\%$ of the total $t\bar{t}$ rate at the LHC. 
Therefore, electroweak production of top quark pairs is not observable at the LHC and does not yield sufficient sensitivity for the study of electroweak couplings. 
Even tree level interference between strong and electroweak amplitudes at  $\mathcal{O}(\alpha \alpha_s)$ is too small: 
It only exists for the parton luminosity suppressed process $b\bar{b} \to t\bar{t}$ process~\cite{Bernreuther:2008md}, which yields two permille of the total rate. 
Fortunately, there is the possibility to access $\mathcal{O}(\alpha)$ effects at higher orders in perturbation theory. 
At next-to-leading order they are the  $\mathcal{O}(\alpha \alpha_s^2 )$ corrections, which comprise electroweak one-loop corrections to the QCD induced process and 
the corresponding single real gluon emission contributions to QCD-electroweak mixed amplitudes (see e.g., Refs.~\cite{Kuhn:2005it,Kuhn:2006vh}). 
Moreover, it is well-known that one-loop diagrams with virtual $Z$ and $W$ bosons are enhanced by logarithms of $\hat{s}\big/M_W^2$, 
which dominate the electroweak correction and enhance sensitivity at high energies. 
Within the SM these corrections have been known for a long time~\cite{Beenakker:1993yr,Kuhn:2005it,Bernreuther:2005is,Bernreuther:2006vg,Kuhn:2006vh,Moretti:2006nf,Kuhn:2013zoa} and 
feature negative corrections that indeed grow with energy. 
In fact, close to the production threshold the corrections are approximately $-2\%$ and they decrease to $-20\%$ at 
top quark transverse momenta of about $2~\mathrm{TeV}$~\cite{Kuhn:2006vh}.
Real emission corrections with $Z$ and $W$ bosons are typically neglected because they lead to a different final state and their unresolved contribution is small~\cite{Baur:2006sn}.
Pure photonic corrections have been calculated in Refs.~\cite{Hollik:2007sw,Pagani:2016caq} and are found to be constantly small at the level of $\mathcal{O}(1\%)$ over a wide range of energy.
\\

We extend the existing results in the literature and calculate, for the first time,  the weak corrections within a SM Effective Field Theory, 
accounting for the contributions of the dimension-six operators in Eqs.~(\ref{eq:EFTop1}--\ref{eq:EFTop3}). 
Hence, we allow for arbitrary deviations from the SM top quark weak interactions that are consistent with the symmetries of the SM. 
Our calculation is of  $\mathcal{O}(\alpha \alpha_s^2 )$ and for 
completeness, we also include the above mentioned suppressed tree level contributions of  $\mathcal{O}(\alpha^2 )$ and  $\mathcal{O}(\alpha \alpha_s )$.
We neglect all higher-order contributions that are suppressed even stronger.
External fields are renormalized in the on-shell scheme, and there is no coupling renormalization, nor effects of operator mixing at the given order.
Our set of input parameters is $M_Z$, $M_W$, $m_t$ and $G_\mathrm{F}$. 

For both, the $q\bar{q}$ and $gg$ initial state,  we calculate final state corrections to $s$-channel diagrams as shown in Fig.~(\ref{fig:svert}), 
using the anomalous couplings in Eqs.~(\ref{eq:ttzcoupl}), (\ref{eq:tbwcoupl}) and (\ref{eq:dGB1}--\ref{eq:dGB4}).
The $gg$ inital state receives corrections to the $t$-channel diagrams in Fig~\ref{fig:ggct}, as well. 
%
\begin{figure}[t]
\centering
\begin{minipage}{0.45\textwidth}
\centering
\raisebox{-0.4\totalheight}{\includegraphics[width=.35\textwidth]{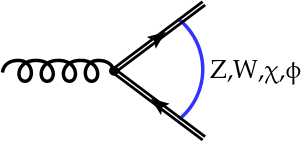}}
{$+$}
\raisebox{-0.4\totalheight}{\includegraphics[width=.23\textwidth]{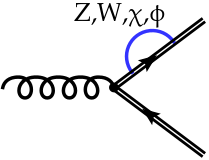}}
{$+$}
\raisebox{-0.5\totalheight}{\includegraphics[width=.23\textwidth]{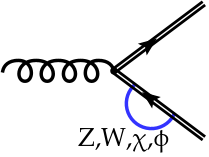}}
\caption{\label{fig:svert} Final state corrections with EFT contributions to the $s$-channel $q\bar{q}$ and $gg$ processes.}
\end{minipage}
\hfill
\begin{minipage}{0.52\textwidth}
\centering
\raisebox{-0.4\totalheight}{\includegraphics[width=.18\textwidth]{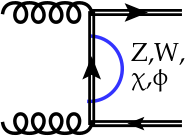}}
{$+$}
\raisebox{-0.4\totalheight}{\includegraphics[width=.18\textwidth]{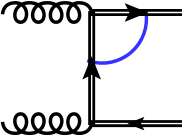}}
{$+$}
\raisebox{-0.4\totalheight}{\includegraphics[width=.18\textwidth]{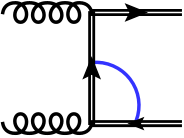}}
{$+$}
\raisebox{-0.4\totalheight}{\includegraphics[width=.18\textwidth]{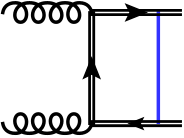}}
\\[4ex]
\raisebox{-0.4\totalheight}{\includegraphics[width=.18\textwidth]{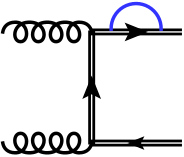}}
{$+$}
{
\raisebox{-0.5\totalheight}{\includegraphics[width=.18\textwidth]{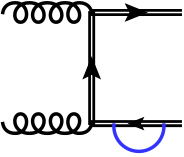}}
}
{$+$}
\raisebox{-0.45\totalheight}{\includegraphics[width=.18\textwidth]{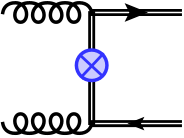}}
\caption{\label{fig:ggct}Loop diagrams and mass counter term receiving EFT contributions in the $t$-channel  $gg$ process.}
\end{minipage}
\end{figure}
%
Loop amplitudes with would-be Goldstone boson exchange increase the integral tensor rank by up to two due to the additional $\slashed{p}$ terms in 
the EFT contributions in Eqs.~(\ref{eq:dGB1}--\ref{eq:dGB4}). 
For example, the first diagram in Fig.~\ref{fig:svert} has a contribution of the form 
\bea
\label{eq:tensorint}
  \sim \frac{v^2}{\Lambda^2} \; 
  \int \! \frac{\mathrm{d}^{D}\ell}{4 \pi} \;
  \frac{ \bar{u}_3 \, \slashed{\ell} ( \slashed{\ell} +\slashed{p}_3 + m_t ) \gamma^\mu (\slashed{\ell} -\slashed{p}_4 + m_t) \slashed{\ell} \, v_4}
  {D_0 D_1 D_2},
\eea
where $D_i = (\ell+k_i)^2 - m_i^2$ and the $k_i$ are linear combinations of the external momenta. This expression contains a rank-four three point integral. 
However, in this case the tensor rank can be lowered by two ranks using
\bea 
\label{eq:reductiontrick}
\slashed{\ell} \slashed{\ell}=D_0+m_0^2\quad \Rightarrow\quad  \frac{\slashed{\ell} \slashed{\ell}}{D_0 D_1 D_2} = \frac1{D_1 D_2} + \frac{m_0^2}{D_0 D_1 D_2}.
\eea
We can reduce all other higher-rank tensor integrals arising from the EFT contributions using Eq.~(\ref{eq:reductiontrick}) and 
arrive at the set of scalar integrals required for the SM calculation. 
We note, however, that additional UV poles appear since the higher-rank tensor integrals yield additional divergencies. 
These UV poles are canceled by a proper renormalization within the effective field theory. 
This is achieved by wave function and mass renormalization which accounts for the extended SMEFT Lagrangian.

Following Ref.~\cite{Denner:1991kt}, the renormalization constants for the bare top quark field $t_0=(1+\frac{1}{2}\delta Z_t)t$ and the bare top quark mass  $m_{t,0}=m_t+\delta m_t$ are given by
\bea
\delta Z_t &=& - \mathrm{Re}\!\left[\Sigma^\mathrm{L}_t(m_t^2) +\Sigma^\mathrm{R}_t(m_t^2)\right]
             - 2 m_t^2 \frac{\partial}{\partial p^2} \mathrm{Re}\!\left[\Sigma^\mathrm{L}_t(p^2)+ \Sigma^\mathrm{R}_t(p^2) + 2 \Sigma^\mathrm{S}_t(p^2) \right] \bigg|_{p^2=m_t^2}{\hspace{-0.5ex},}\hspace{5.5ex}
\\
\delta m_t &=& \frac{m_t}{2} \mathrm{Re}\!\left[ \Sigma^\mathrm{L}_t(m_t^2)+ \Sigma^\mathrm{R}_t(m_t^2) + 2 \Sigma^\mathrm{S}_t(m_t^2) \right],
\eea
where the chiral top quark self energies $\Sigma^\lambda_t(p^2)$ are evaluated at $\mathcal{O}(\alpha)$ including 
contributions from the EFT operators in Eqs.~(\ref{eq:EFTop1}--\ref{eq:EFTop3}) for $\lambda=\mathrm{L,R,S}$~\cite{Denner:1991kt}.
We write the self energy as a sum over all particle contributions $\Sigma^\lambda_t(p^2) = \sum_{\varphi} \Sigma^\lambda_{t\varphi}(p^2)$
where $\varphi$ runs over $Z,W,\chi,\phi$ and $H$. 
For the contributions receiving EFT modifications we find  
\bea
\Sigma^\lambda_{tZ}(p^2) &=& \Sigma^{\lambda, \mathrm{SM}}_{tZ}(p^2) \bigg|_{\text{replacement of Eq.~(\ref{eq:ttzcoupl})}},\label{eq:selfenf}
\\
\Sigma^\lambda_{tW}(p^2) &=& \Sigma^{\lambda, \mathrm{SM}}_{tW}(p^2) \bigg|_{\text{replacement of Eq.~(\ref{eq:tbwcoupl})}},
\\
\Sigma^\mathrm{L}_{t\chi}(p^2) &=& -\frac{\alpha}{4\pi} \frac1{4 s_\mathrm{w}^2 c_\mathrm{w}^2 M_Z^2} \bigg\{ 
m_t^2 B_1(p^2;m_t^2,M_Z^2) +  4 m_t^2 \bigg( B_0(p^2;m_t^2,M_Z^2) 
\nonumber \\
&&+ B_1(p^2;m_t^2,M_Z^2) \bigg) \frac{v^2}{\Lambda^2} C_{33}^{\varphi q 3} + \mathcal{O}\left(\frac{v^4}{\Lambda^4}\right)
\bigg\},
\eea
\bea
\Sigma^\mathrm{R}_{t\chi}(p^2) &=& -\frac{\alpha}{4\pi} \frac1{4 s_\mathrm{w}^2 c_\mathrm{w}^2 M_Z^2} \bigg\{ 
m_t^2 B_1(p^2;m_t^2,M_Z^2) +  2 m_t^2 \bigg( B_0(p^2;m_t^2,M_Z^2) 
\nonumber \\ 
&&+ B_1(p^2;m_t^2,M_Z^2) \bigg) \frac{v^2}{\Lambda^2} C_{33}^{\varphi u} + \mathcal{O}\left(\frac{v^4}{\Lambda^4}\right)
\bigg\},
\\
\Sigma^\mathrm{S}_{t\chi}(p^2) &=& -\frac{\alpha}{4\pi} \frac1{4 s_\mathrm{w}^2 c_\mathrm{w}^2 M_Z^2} \bigg\{ 
m_t^2 B_0(p^2;m_t^2,M_Z^2) +  m_t^2 \bigg( B_0(p^2;m_t^2,M_Z^2) 
\nonumber \\
&& + \frac{p^2}{m_t^2} B_1(p^2;m_t^2,M_Z^2) + \frac{A_0(M_Z^2)}{m_t^2} \bigg) \frac{v^2}{\Lambda^2} \left(2 C_{33}^{\varphi q 3} + C_{33}^{\varphi u} \right) 
+ \mathcal{O}\left(\frac{v^4}{\Lambda^4}\right)
\bigg\},\hspace{3ex}
\\
\Sigma^\mathrm{L}_{t\phi}(p^2) &=& \mathcal{O}\left(\frac{v^4}{\Lambda^4}\right),
\\
\Sigma^\mathrm{S}_{t\phi}(p^2) &=& -\frac{\alpha}{4\pi} \frac1{2 s_\mathrm{w}^2 c_\mathrm{w}^2 M_Z^2} 
\bigg( p^2 B_1(p^2;0,M_W^2) + A_0(M_W^2) \bigg) \frac{v^2}{\Lambda^2} C_{33}^{\varphi q 3}. \label{eq:selfenl}
\eea
For brevity we only show results up to $\mathcal{O}({v^2}\big/{\Lambda^2})$ here. 
However, our simulation (optionally) includes the $\mathcal{O}({v^4}\big/{\Lambda^4})$ contribution from squared dimension-six operators. Note that contributions not listed in Eqs.~(\ref{eq:selfenf}--\ref{eq:selfenl}) retain their SM expressions without modifications from the EFT operators.

The UV finite box diagram contributions, shown in Fig.~\ref{fig:boxes}, contain additional infrared (IR) poles from soft and collinear gluons in the loop.
They are canceled by mixed QCD-electroweak amplitudes with real gluon emission, which are also sensitive to the EFT operators. 
Hence, we consistently include them in our analysis and confirm a cancellation of all IR poles for arbitrary top quark couplings.

\subsection{Benchmark process $pp \to t\bar{t}+Z$}

To benchmark our results obtained from $pp \to t\bar{t}$ we also consider the $pp \to t\bar{t}+Z$ process, 
which yields sensitivity to the same kind of EFT operators in Eqs.~(\ref{eq:EFTop1}--\ref{eq:EFTop3}).
In contrast to the virtual corrections in $t\bar{t}$ production, $\mathcal{O}(\alpha)$ dependence arises directly from the on-shell coupling of the $Z$ boson to the top quarks at tree level. 
In the case that $Z\to \ell\ell$ the final state is a clean experimental signature with relatively small backgrounds. 
On the theory side, higher-order predictions such as the NLO corrections to the SM process have been known for many years.
In this work we build upon the \textsc{TOPAZ} framework%
\footnote{The code is available at \href{https://github.com/TOPAZdevelop/TOPAZ}{\texttt{https://github.com/TOPAZdevelop/TOPAZ}}}~\cite{Topaz:2014}, which was used in Refs.~\cite{Rontsch:2014cca,Rontsch:2015una,Schulze:2016qas} to calculate the NLO QCD corrections to $t\bar{t}+Z$ production and decay including anomalous couplings between the top quark and the $Z$ boson. 
These anomalous couplings can be related directly  to the EFT coefficients in Eq.~(\ref{eq:ttzcoupl}) and allow an immediate comparison with 
the $t\bar{t}$ process.

%
%
\begin{figure}[t]
\centering
\raisebox{-0.4\totalheight}{\includegraphics[width=.15\textwidth]{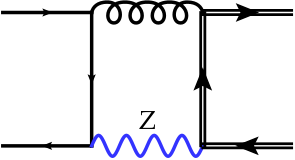}}
\hspace{1ex}
{$+$}
\hspace{1ex}
\raisebox{-0.4\totalheight}{\includegraphics[width=.15\textwidth]{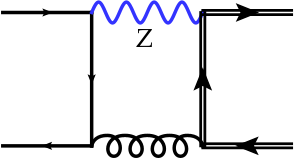}}
\hspace{2.5cm}
\raisebox{-0.4\totalheight}{\includegraphics[width=.15\textwidth]{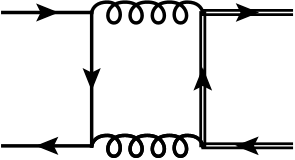}}
{$\times$}
\raisebox{-0.4\totalheight}{\includegraphics[width=.15\textwidth]{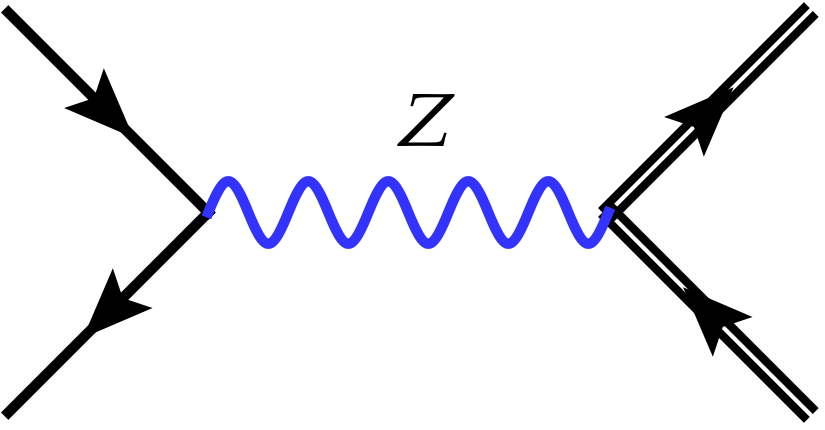}}
\hspace{-1mm}\raisebox{+4\height}{$*$}
\\[4ex]
\raisebox{-0.4\totalheight}{\includegraphics[width=.16\textwidth]{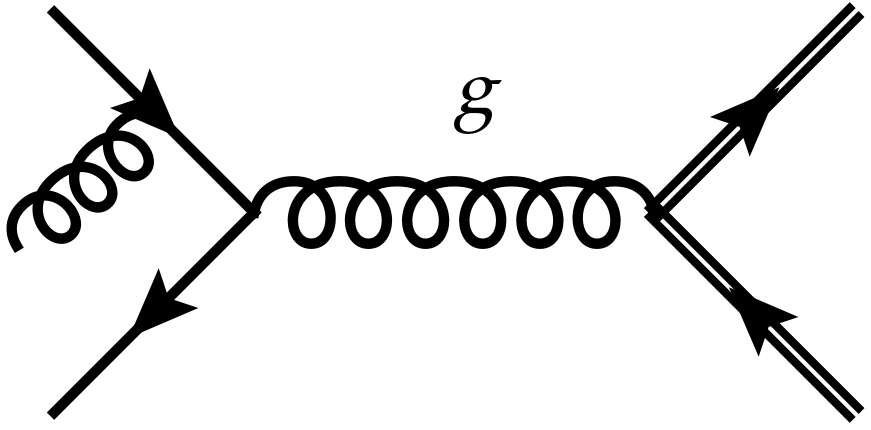}}
{$\times$}
\raisebox{-0.4\totalheight}{\includegraphics[width=.16\textwidth]{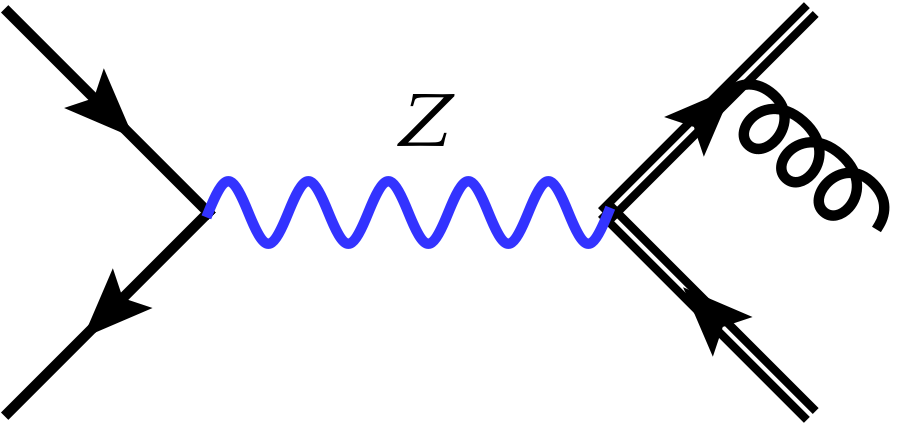}}
\hspace{-3mm}\raisebox{+3.5\height}{$*$}
\caption{\label{fig:boxes} 
Diagrams and diagram interferences containing infrared singularities from soft and collinear gluons in the $q\bar{q}$ initial state. 
}
\end{figure}
%

\subsection{Numerical evaluation}

We implement the analytic results of our calculation in \textsc{Fortran} and make it available as an external add-on to the Monte-Carlo simulator \textsc{MCFM}~\cite{Campbell:2015qma}.
\textsc{MCFM} allows us to benefit from its easy-accessible, well-tested and established simulation framework. 
Moreover, the electroweak correction in the SM are already incorporated~\cite{Campbell:2016dks} and allow us to do valuable cross checks. 
Of course, we fully reproduce the \textsc{MCFM} results when setting $C_{33}^{\varphi q 3}= C_{33}^{\varphi u}=0$ in our calculation.
Our add-on can be incorporated into the publicly available version by simply replacing eight files and re-compiling the code%
\footnote{The files are available at \href{https://github.com/TOPAZdevelop/MCFM-8.3_EWSMEFT_ADDON}{\texttt{https://github.com/TOPAZdevelop/MCFM-8.3\_EWSMEFT\_ADDON}}. Further instructions are contained in the \texttt{README} file.}. 
The numerical values for our input parameter are
\bea
M_Z=91.1876\,\mathrm{GeV}, \quad 
M_W=80.385\,\mathrm{GeV}, \quad 
M_H=125.0\,\mathrm{GeV}, \quad 
\\
G_\mathrm{F} = 1.16639 \times 10^{-5} \,\mathrm{GeV}^{-2},\quad 
m_t=173.2\,\mathrm{GeV}, \quad 
m_b=0.0\,\mathrm{GeV}.
\eea
We use the parton distribution functions \texttt{NNPDF30\_nnlo\_as\_0118\_qed}~\cite{Ball:2014uwa} and its corresponding value of $\alpha_s$.
For simplicity, the top quarks are simulated as stable particles in our calculations. 
We therefore multiply the cross sections with the branching fraction of the {\it lepton+jet} 
final state $\mathrm{Br}(\ell+\mathrm{jets})=8/27\approx 30\%$ and allow electrons and muons in the final state. 
This final state allows an unambiguous reconstruction of the top quark momenta and justifies the 
approximation of stable top quarks. 
To account for the state-of-the-art QCD predictions for $t\bar{t}$ production in an approximate way, we match the scale setting of Ref.~\cite{Czakon:2017wor} 
$\mu=m_{\mathrm{T},t}/2$ and multiply our weak corrections by the NNLO QCD $K$-factor of $1.67$~\cite{Czakon:2011xx}.
This corresponds to the so-called {\it multiplicative approach}, which was found to be preferred over an additive approach~\cite{Czakon:2017wor}.
The respective QCD scale uncertainty of ${}_{-6\,\%}^{+4\,\%}$ is taken from Ref.~\cite{Czakon:2011xx} and set to $\pm 5\,\%$ in our analysis. 
For the $pp \to t\bar{t}+Z$ process, we multiply with a $K$-factor of $1.23$ and assume an uncertainty of $\pm 15\,\%$~\cite{Topaz:2014}. 
The $Z$ boson is assumed to decay into either electron or muon pairs. 
\\

For estimating the sensitivity to New Physics in our processes, we need to find the functional dependence of the production rate 
on the EFT coefficients $C_{33}^{\varphi q 3}$ and $C_{33}^{\varphi u}$. 
Since we also want to exploit the effects from electroweak Sudakov logarithms that grow with energy, we even need to find this dependence for 
differential distributions. 
If we truncate all terms beyond $\mathcal{O}(v^4\big/\Lambda^4)$, we can parameterize each bin $i$ of a differential distribution as 
\bea
\label{eq:binfit}
     N_i^\textrm{SMEFT} = {n_0}_i+{n_1}_i\;C^{\varphi q3}_{33}+{n_2}_i\;C^{\varphi u}_{33} 
     +{n_3}_i\;(C^{\varphi q3}_{33})^2+{n_4}_i\;(C^{\varphi u}_{33})^2+{n_5}_i\;C^{\varphi q3}_{33} \, C^{\varphi u}_{33}, \quad \quad 
\eea
with coefficients ${n_j}_i$. 
These coefficients can be determined by generating six differential distributions for six different values of $C^{\varphi q3}_{33}$ and $C^{\varphi u}_{33}$. 
Once this fit is obtained, it can be used to predict the differential distribution for \textit{any} combination of the EFT coefficients
and it can even be truncated at $\mathcal{O}(v^2\big/\Lambda^2)$ in order to remove dimension-eight terms from squared dimension-six terms. If not stated otherwise, we keep the squared dimension-six terms in the SMEFT predictions in this work.
Using the parameterization in Eq.~(\ref{eq:binfit}) we perform Pearson-$\chi^2$ hypothesis tests
\bea
\label{eq:chi2}
  \chi^2(C^{\varphi q3}_{33},C^{\varphi u}_{33})=\sum\limits_{i}
  \frac{\left( N_i^\textrm{SMEFT} - N_i^\textrm{SM}\right)^2}{N_i^\textrm{SMEFT}}
\eea
on the histogrammed distributions with $N^\textrm{SM}_i={n_0}_i$, where the sum runs over all bins with $N_i\geq 5$.
The null hypothesis is that the observed data follows the expected prediction. 
In the following, we assume that the observed data is given by the SM prediction and the expected result is from the SMEFT.
This allows us to quote the $p$-value for the case that the SM result is a fluctuation of the SMEFT distribution, 
which can be translated into a $95\,\%$ confidence level (C.L.) exclusion for the given hypothesis. 
To account for systematic uncertainties $\Delta_{\textrm{sys.}}$ from scale variation we rescale the $N_i^H \to f_H \cdot N_i^H$ with factors 
\begin{equation}
\label{eq:rescale}
  (1-\Delta_{\textrm{sys.}})\leq f_H\leq(1+\Delta_{\textrm{sys.}}),
\end{equation}
such that $|f_\textrm{SMEFT}\sum\limits_iN^{\textrm{SMEFT}}_i-f_\textrm{SM}\sum\limits_iN^{\textrm{SM}}_i|$ is minimized.
This brings the total cross sections of two hypotheses closest while maintaining their shape differences.

\section{Results}

\begin{figure}[t]
\centering
\begin{subfigure}{0.49\textwidth}
\centering
\includegraphics[height=0.33\textheight]{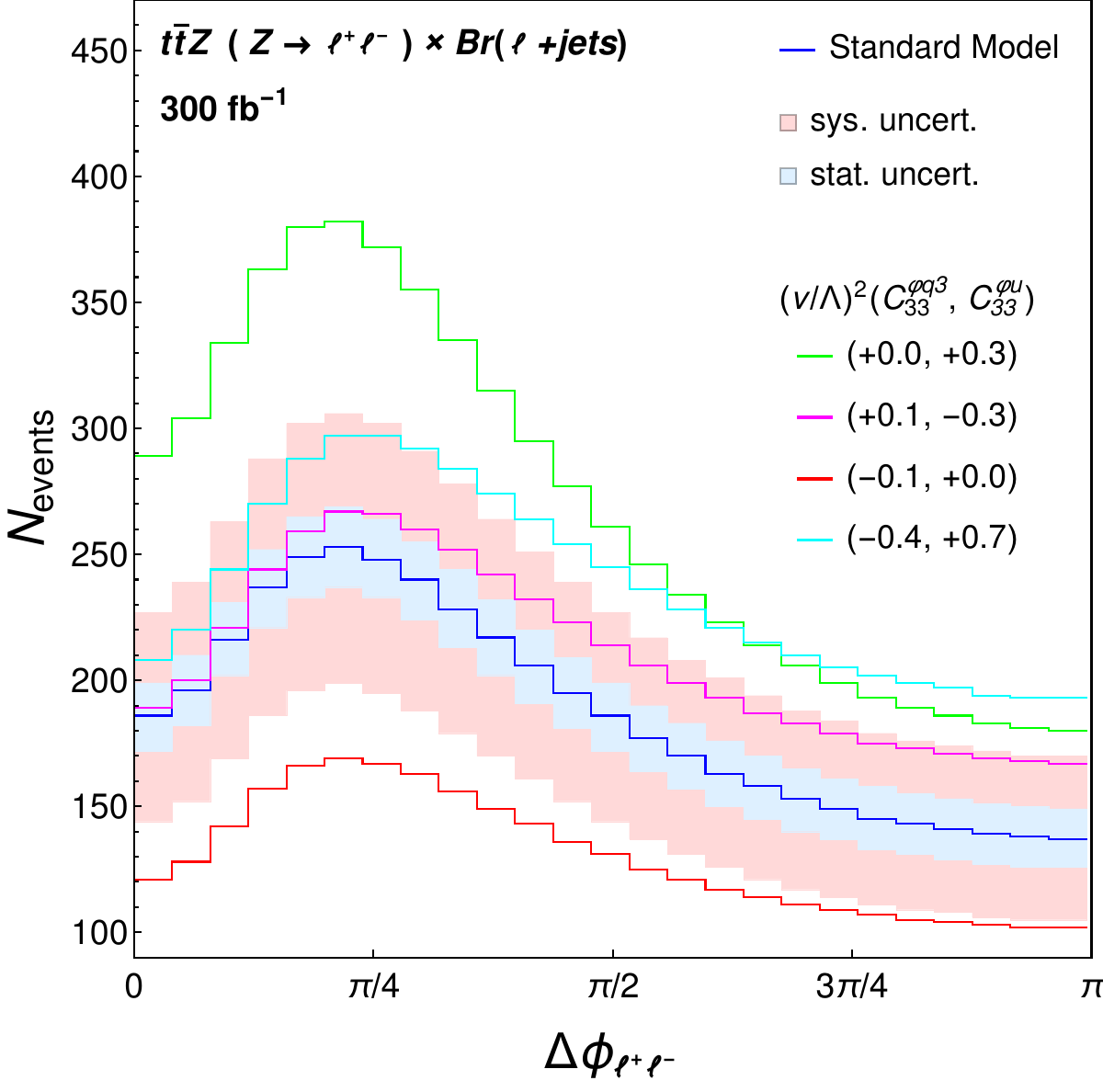}
\end{subfigure}
\hfill
\begin{subfigure}{0.49\textwidth}
\centering
\includegraphics[height=0.33\textheight]{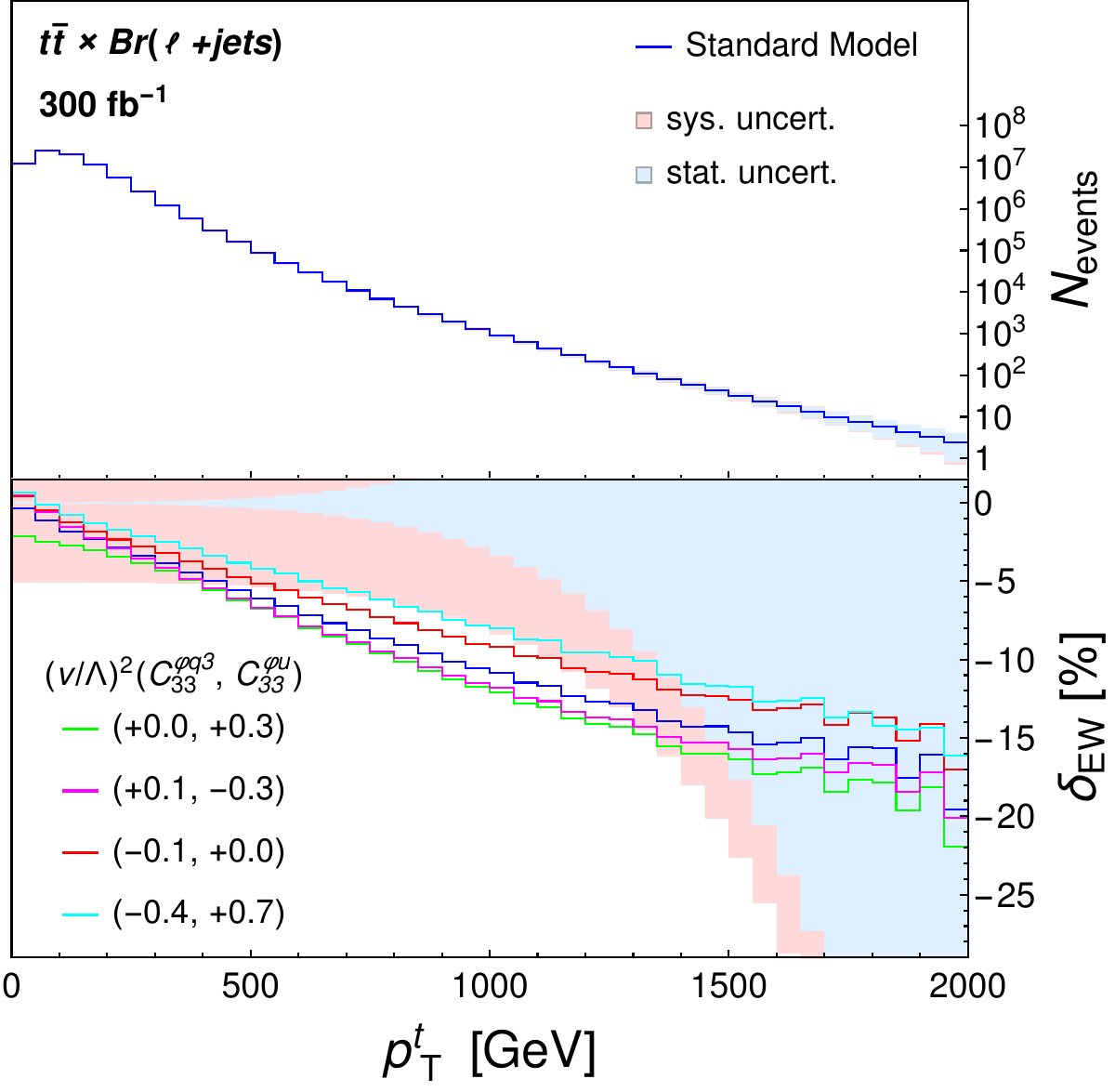}
\end{subfigure}
\caption{\label{fig:ttdelptt}
Left: Distribution of $\Delta\phi_{\ell\ell}$ in $t\bar{t}+Z$ production at 13 TeV. 
Right: $p_\mathrm{T}^t$ distribution in $t\bar{t}$ production at 13 TeV and the relative weak correction below. 
The SM prediction is shown in blue, together with statistical and systematic uncertainty bands. 
Modifications from SMEFT contributions are shown for four selected values of $C^{\varphi q3}_{33}$ and $C^{\varphi u}_{33}$.}
\end{figure}

In the following, we present the phenomenological results of our calculation for the $13$~TeV LHC, assuming an integrated luminosity of $300\,\ifb$.
For a direct comparison we present our findings for $t\bar{t}$ and $t\bar{t}+Z$ side by side. 
We choose kinematic variables that show strong sensitivity to New Physics for each of the respective processes. 
In the case of $t\bar{t}+Z$ production the azimuthal opening angle of the two leptons $\Delta\phi_{\ell\ell}$ from the $Z$ boson decay was identified as the
most sensitive observable in Refs.~\cite{Rontsch:2014cca,Rontsch:2015una}. 
For $t\bar{t}$ production we choose the transverse momentum $p_\mathrm{T}^t$ of the top quark because it features the largest 
electroweak correction for a given energy scale. 
In Fig.~\ref{fig:ttdelptt} we show their distributions for the SM prediction (blue line), including the statistical uncertainties and the 
systematic uncertainties from QCD scale variation. 
For the steeply falling $p_\mathrm{T}^t$ distribution we also show the relative $\mathcal{O}(\alpha)$ correction with respect to the leading-order prediction,
which emphasizes the electroweak Sudakov enhancement (with negative sign) of corrections growing with energy. 
At first glance, the  significantly different numbers of events between the two processes stick out:
While the $t\bar{t}+Z$ histogram is saturated with $\mathcal{O}(100)$ events per bin, the distribution for $t\bar{t}$ peaks with more than $10^7$ at lower $p_\mathrm{T}^t$. 
Evidently this directly impacts the respective statistical uncertainties and it is one of the major impact factors on the New Physics constraints presented below. 
Systematic uncertainties also differ because $t\bar{t}+Z$ production is currently known at NLO QCD whereas $t\bar{t}$ is known at NNLO QCD.
In Fig.~\ref{fig:ttdelptt} we also present four selected New Physics predictions for some values of EFT coefficients.
The differences in both normalization and shape with respect to the SM prediction are clearly visible and suggest good sensitivity to New Physics.

\begin{figure}[t]
\centering
\begin{subfigure}{0.49\textwidth}
\centering
\includegraphics[height=0.33\textheight]{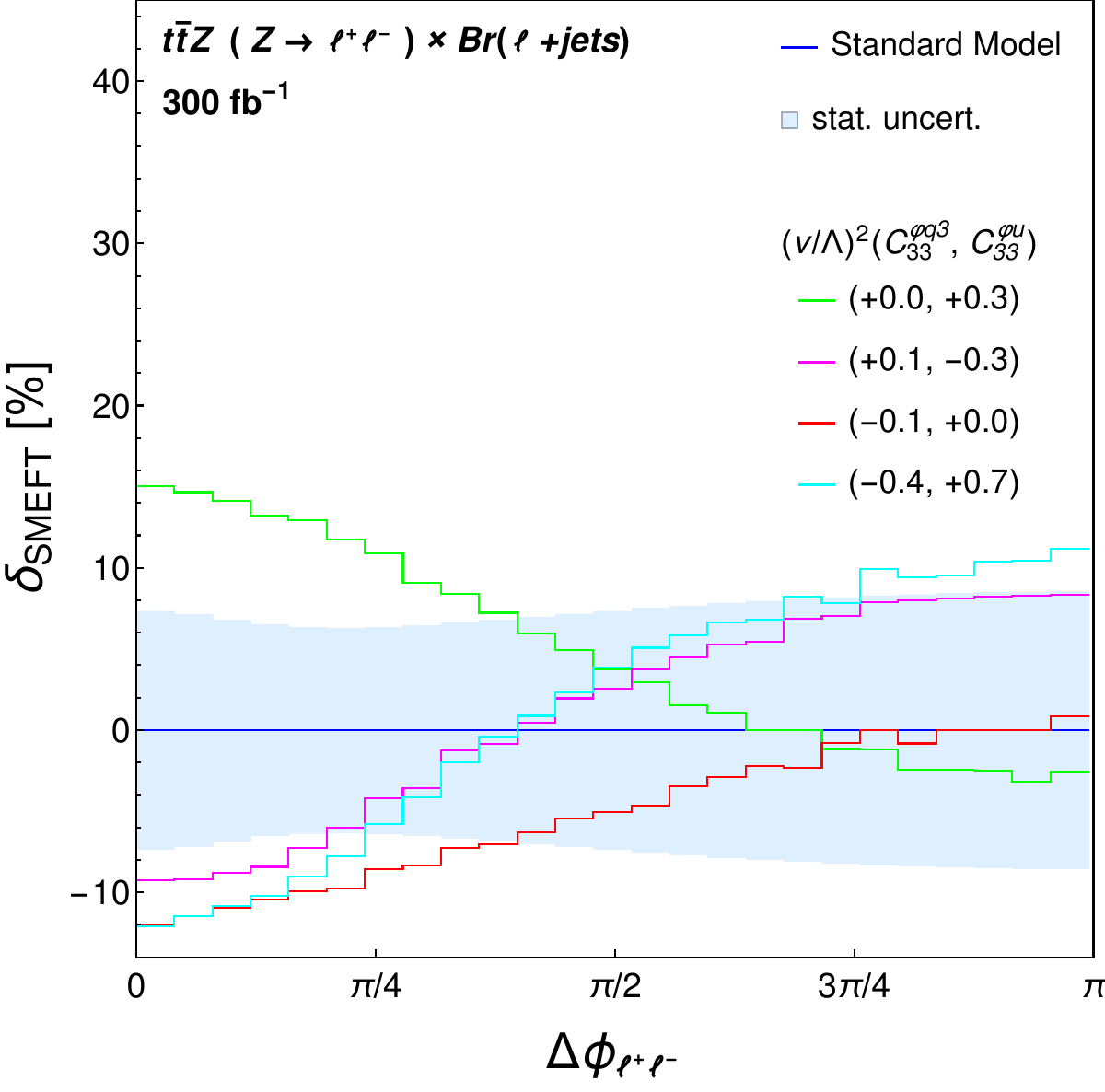}
\end{subfigure}
\hfill
\begin{subfigure}{0.49\textwidth}
\centering
\includegraphics[height=0.33\textheight]{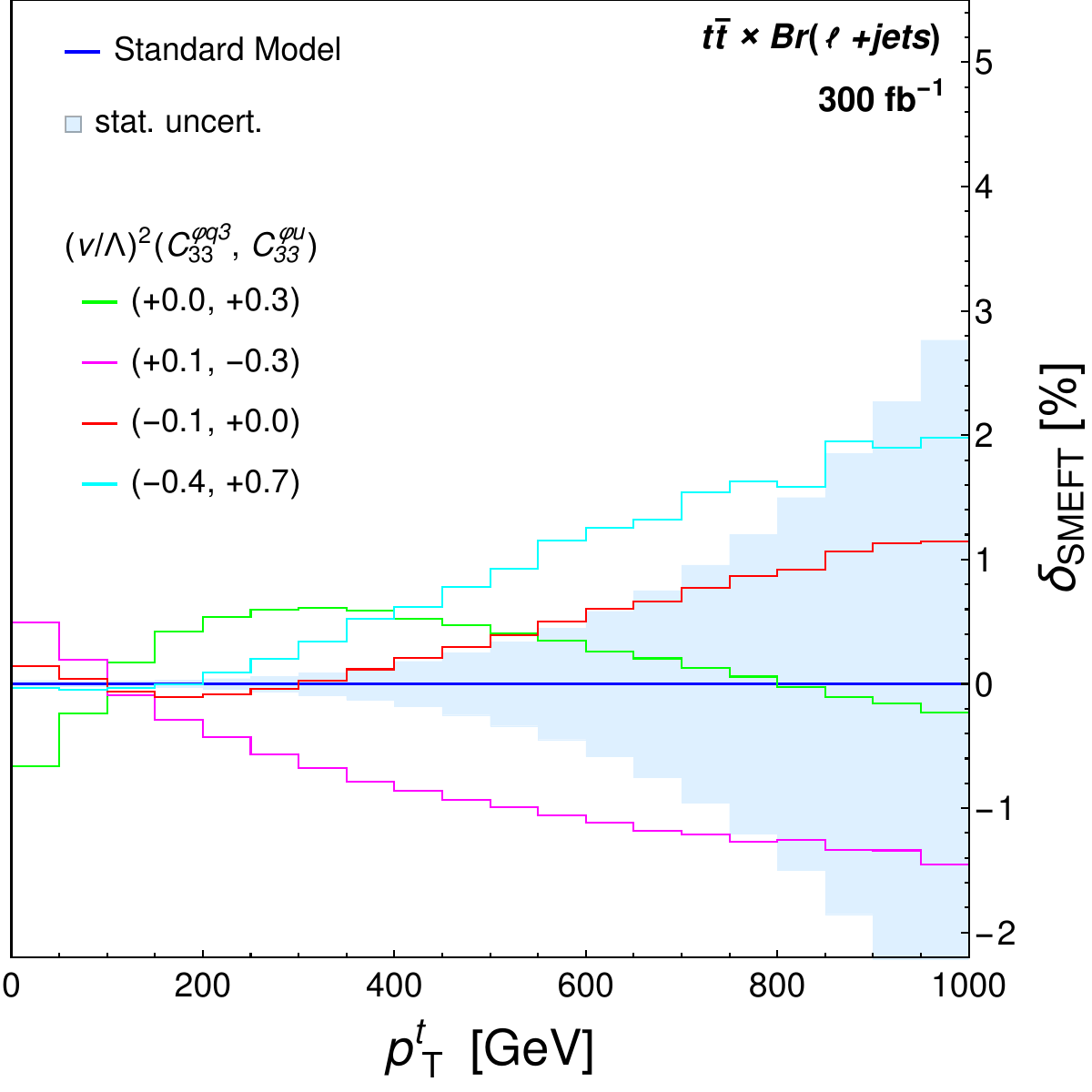}
\end{subfigure}
\caption{\label{fig:relSMEFT}
Size of the relative modifications due to SMEFT contributions (green, magenta, red, and cyan) with respect to Standard Model expectation (blue), 
after rescaling to account for systematic uncertainties. The light blue band shows statistical uncertainties. }
\end{figure}
Before quantifying the exclusion power on the EFT coefficients, we study the origin of this sensitivity in more detail. 
Therefore, in Fig.~\ref{fig:relSMEFT} we plot the relative SMEFT correction with respect to the NLO SM prediction, after 
accounting for systematic uncertainties through the rescaling described around Eq.~(\ref{eq:rescale})
\bea
\delta_{\textrm{SMEFT},i}=\frac{N_i^\textrm{SMEFT}}{N_i^\textrm{SM}}-1.
\eea
The $\Delta \phi_{\ell\ell}$ distribution (left) of $t\bar{t}+Z$ features shape variations between $-10\,\%$ and $+15\,\%$ depending on the SMEFT parameter point
and show  almost constant uncertainty of about $\pm 8\%$ in the whole range.
Hence, we roughly expect between $1$--$2\,\sigma$ C.L. exclusion limits for these points. 
For the same SMEFT parameter points, the $p_\mathrm{T}^t$ distribution in $t\bar{t}$ production shows shape changes that are significantly 
larger than the statistical uncertainties over a wide energy range. 
Only beyond $p_\mathrm{T}^t \approx 600\,\mathrm{GeV}$, there are less and less events that let statistical uncertainties grow to wash out the sensitivity. 
This is a first hint that we can expect strong New Physics constraints from the electroweak loop corrections to $t\bar{t}$.

Let us now come to our main results. 
Using the fit in Eq.~(\ref{eq:binfit}) for the kinematic variables discussed above, 
we scan a wide two-dimensional parameter space for the EFT coefficients $C^{\varphi q3}_{33}$ and $C^{\varphi u}_{33}$
and perform a Pearson-$\chi^2$ hypothesis test on the histograms presented in Fig.~\ref{fig:relSMEFT}. 
This allows us to quote exclusion contours for a given set of SMEFT parameter values.
In the left plot in Fig.~\ref{fig:tt_95clvol4} we show two such exclusion contours as a function of the $(v/\Lambda)^2 \cdot C_\mathrm{EFT}$
for $t\bar{t}$ production at the 13 TeV LHC with $300\,\ifb$.
The orange contour corresponds to the strict expansion up to dimension-six, whereas the yellow contour contains the squared 
dimension-six terms of order $(v/\Lambda)^4$. 
Apart from the lower right edge, the differences are minimal, suggesting sufficient convergence of the EFT expansion in terms of powers of $\Lambda^{-1}$.
The constraints appear quite strong as the $2$-$\sigma$ exclusion contour ranges only between $\pm\, 0.1$\footnote{In appendix~\ref{app:A} the scaling of the exclusion contour with tighter and tighter cuts on the transverse momentum of the top quark is shown.}.
The right plot of Fig.~\ref{fig:tt_95clvol4} contrasts this exclusion limit with the one from $t\bar{t}+Z$ production%
\footnote{We note that the exclusion contour for $t\bar{t}+Z$ production matches very well the one in Ref.~\cite{Rontsch:2014cca}, although 
a different input parameter set and a different statistical analysis is used. }%
, where we keep dimension-eight contributions in both predictions and apply the same statistical analysis. 
This direct comparison shows the strikingly stronger limits from $t\bar{t}$ production. 
For example, the four selected points in Fig.~\ref{fig:tt_95clvol4}~(right) are lying inside the $2\,\sigma$ C.L. contour
of $t\bar{t}+Z$ production and, therefore, cannot be excluded using this process. 
Contrarily, the same points are well outside the corresponding $t\bar{t}$ contour and can be strongly excluded. 
It is also remarkable that $C_{33}^{\varphi u}$ is so much stronger bounded.
Its allowed range is more than a factor of five smaller in $t\bar{t}$ than in $t\bar{t}+Z$.
This is particularly important because $C_{33}^{\varphi u}$ only modifies the right-handed $ttZ$ coupling $d_\mathrm{R}^Z$ in Eq.~(\ref{eq:ttzcoupl})
and does not affect the $tbW$ vertex. 
Therefore, it cannot be further constrained from single top quark production or top quark decay analyses. Refs.~\cite{Hartland:2019bjb,Brivio:2019ius} report comparatively loose bounds obtained from studying $tZ$ production which also confirm that electroweak corrections to $t\bar{t}$ are apparently the strongest probe available at the LHC. 
\begin{figure}[t]
\centering\begin{subfigure}{0.49\textwidth}
\centering
\includegraphics[height=0.315\textheight]{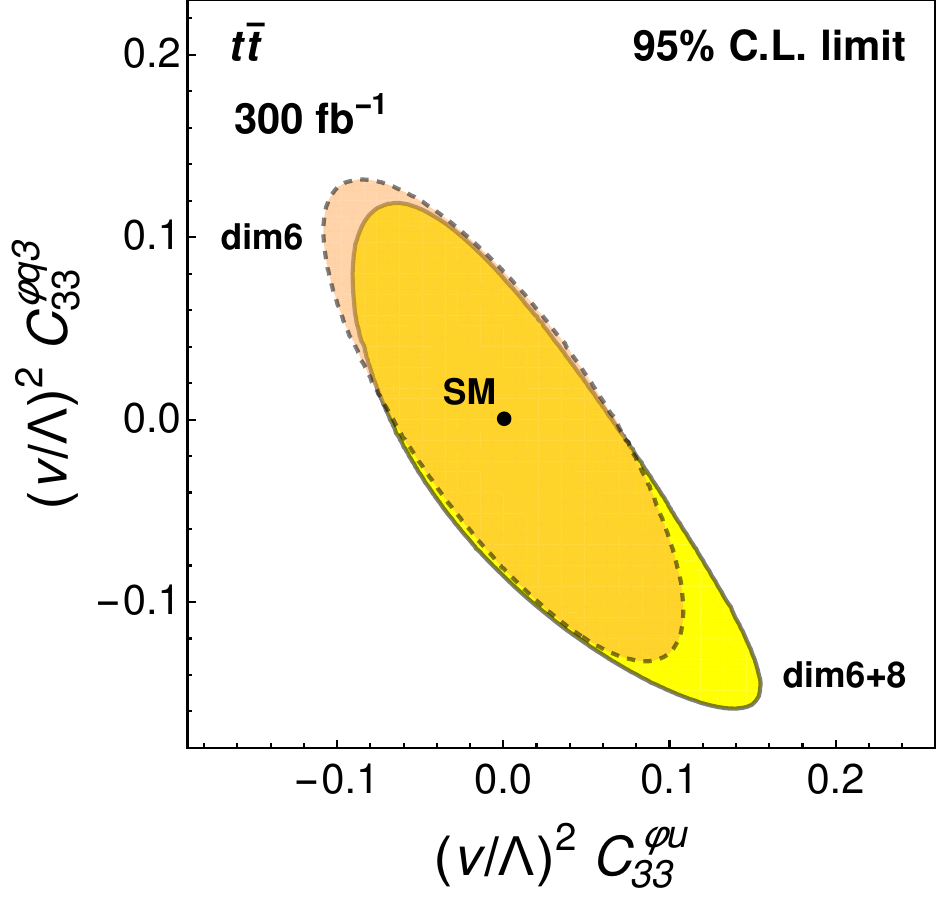}
\end{subfigure}
\hfill
\begin{subfigure}{0.49\textwidth}
\centering
\includegraphics[height=0.315\textheight]{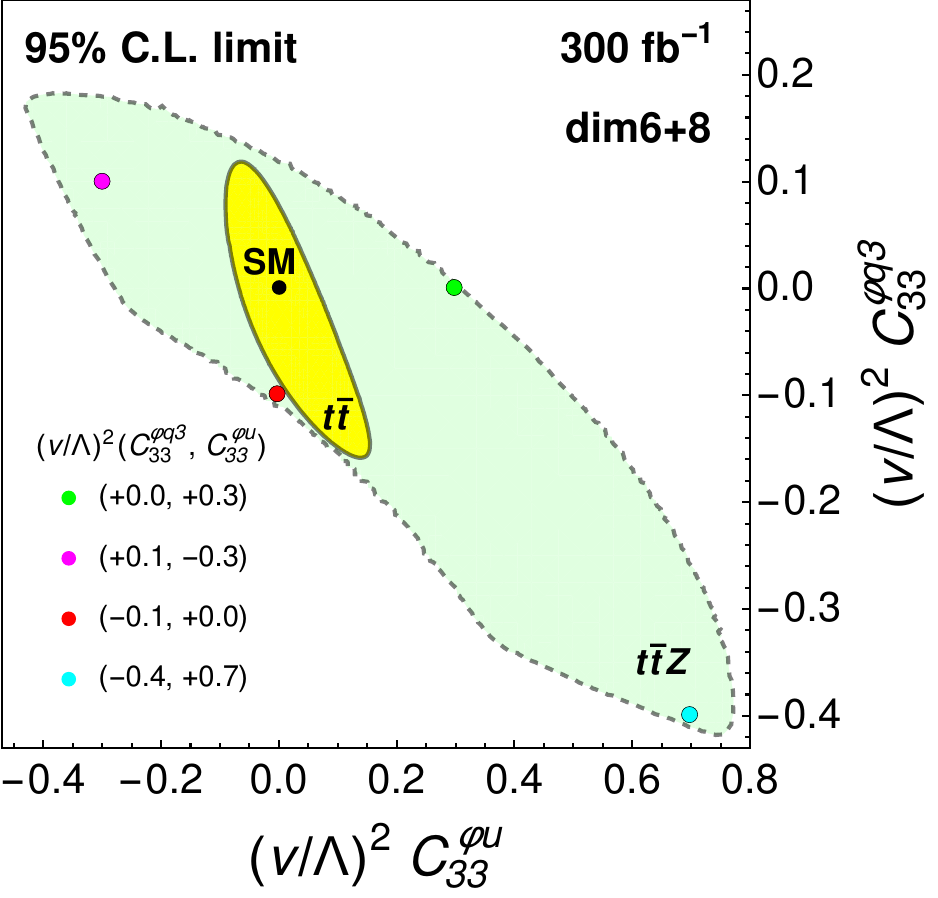}
\end{subfigure}
\caption{\label{fig:tt_95clvol4}Left: Exclusion limits on EFT coefficients including (yellow, solid) and neglecting (orange, dashed) the squared dimension-six terms. 
Right: Exclusion limits obtained from the $p^t_T$ distributions for $t\bar{t}$ production including EW corrections (yellow, solid) and the $\Delta\phi_{\ell^+\ell^-}$ distributions for $t\bar{t}+Z$ production (light green, dashed). Colored dots correspond to particular values of the EFT coefficients.}
\end{figure}

Finally, we compare our projected constraints with the most up-to-date experimental results. 
Members of the CMS experiment recently published an analysis of $t\bar{t}+Z$ using $77.5\,\ifb$ integrated luminosity at 13~TeV~\cite{CMS:2019too}.
Similar to our work here, they assume the decay $Z\to \ell\ell$ with electrons and muons. 
However, they allow three or four charged leptons in the final state such that also di-leptonic $t\bar{t}$ decays are included.
They interpret their results in terms of the same EFT coefficients that we use here 
($c^-_{\varphi Q} \hat{=} C_{33}^{\varphi q 1}-C_{33}^{\varphi q 3} $ and $c_{\varphi t} \hat{=} C_{33}^{\varphi u} $), so 
that a rough comparison is possible. 
If we take Fig.~7 of Ref.~\cite{CMS:2019too} and project the edges of the (upper) 2-dimensional 95~\% C.L. contour onto the axes, we find in our notation 
\bea
   (v/\Lambda)^2\;C_{33}^{\varphi q 3} &\in& \big[-0.46\,...\,0.21\big]_\text{CMS},
    \nonumber \\
    (v/\Lambda)^2\;C_{33}^{\varphi u} &\in&   \big[-0.55\,...\,0.79\big]_\text{CMS}.   
\eea
These numbers are in decent agreement (although somewhat weaker) with our presented results in Fig.~\ref{fig:tt_95clvol4}~(right), 
keeping in mind that we assumed 
a higher integrated luminosity and just the lepton plus jet $t\bar{t}$ decay channel.
Yet, we take this agreement as an indication for the robustness of our analysis and our conservative uncertainty treatment. 
It strengthens our proposal to explore the $t\bar{t}$ final state in future analyses and exploit its superiority.

\section{Conclusions}

In this paper, we investigate the question whether New Physics can be probed through electroweak loop corrections in $t\bar{t}$ production at the LHC
and how it compares to direct on-shell probes such as the $pp\to t\bar{t}+Z $ process.
The answer is not at all obvious because of two competing effects:
The available cross section for top quark pair production is very large and perturbatively under excellent control, allowing for a very precise analysis.
Yet, sensitivity only arises at loop level and through off-shell degrees of freedom. 
For the $t\bar{t}+Z $ process the opposite is true, at the respective coupling order. 

To tackle this question, we calculate the $\mathcal{O}(\alpha)$ weak correction to top quark pair production 
and include New Physics through a Standard Model Effective Field Theory parameterization. 
In particular, we allow for anomalous interactions between top quarks and the weak gauge bosons.
We study differential distributions to exploit enhanced sensitivity from electroweak Sudakov logarithms. 
After carefully accounting for the most dominant statistical and systematic uncertainties, 
we find that the $t\bar{t}$ process at $\mathcal{O}(\alpha\alpha_s^2)$ is significantly more sensitive to New Physics 
than the  $t\bar{t}+Z $ process. 
Hence, virtual loop corrections are prevailing over probes with on-shell degrees of freedom. 
This opens up a promising new way for electroweak top quark studies in $t\bar{t}$ production  at the LHC. 

Our study leaves room for further theoretical investigations and future refinements.
For example, a calculation including top quark decays will improve the physics modeling and yield additional sensitivity to a 
subset of the studied EFT operators here. 
EFT operators affecting the QCD production dynamics or the top quark Yukawa interaction can also be added to extend the current analysis. 
On should also keep in mind that a realistic modeling of experimental aspects such as detector effects and efficiencies 
will, to some extend, weaken the sensitivity.
Yet, a first comparison with real experimental data for $t\bar{t}+Z$ production shows that our projections are solid. 
Therefore, we are looking forward to the first application of our proposal in an experimental analysis of the $pp\to t\bar{t}$ process using real LHC data. 

\acknowledgments
We are grateful to AG PEP at Humboldt-University Berlin for providing the computing resources needed for this work. We thank Marcel Vos for giving constructive feedback on the manuscript.

\newpage
\appendix
\section{Dependence of the exclusion limits on $p^t_T$ cuts}\label{app:A}
In Fig.~\ref{fig:tt_95clptcuts} we illustrate the dependence of the exclusion limits from Fig.~\ref{fig:tt_95clvol4} on cuts on the top quark transverse momentum between $p^t_T<400$ GeV and $p^t_T<100$ GeV. We note that these cuts correspond to upper bounds of the invariant mass of the $t\bar{t}$ system between $m_{t\bar{t}}\lessapprox 870$ GeV and $m_{t\bar{t}}\lessapprox 400$ GeV indicating the maximal reach in energy of the respective measurement.
\begin{figure}[h]
\centering
\includegraphics[height=0.33\textheight]{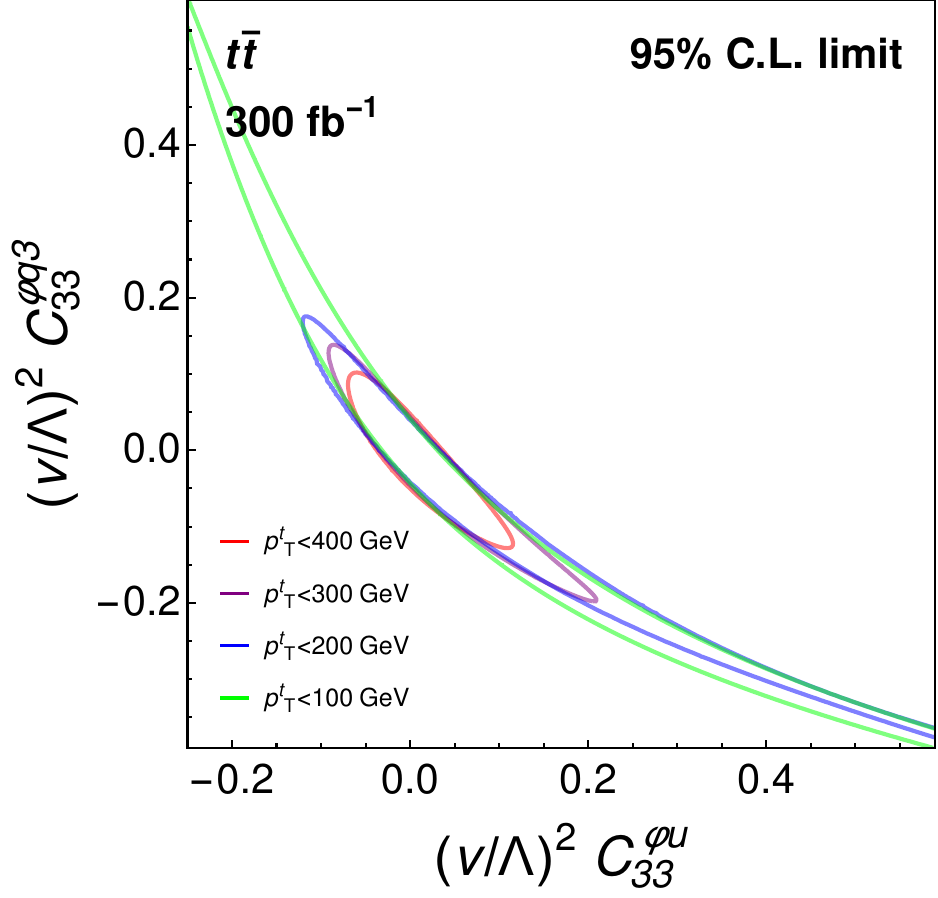}
\caption{\label{fig:tt_95clptcuts}
Exclusion limits obtained from the $p^t_T$ distributions for $t\bar{t}$ production including EW corrections for different cuts on the transverse momentum of the top quark.}
\end{figure}


\providecommand{\href}[2]{#2}\begingroup\raggedright\endgroup

\end{document}